\documentclass[12pt]{article}
\newcommand{\be}{\begin{equation}}  \newcommand{\ee}{\end{equation}}
\newcommand{\bea}{\begin{eqnarray}} \newcommand{\eea}{\end{eqnarray}}
\newcommand{\ba}{\begin{array}}     \newcommand{\ea}{\end{array}}
\newcommand{\ds}{\displaystyle}
\newcommand{\nn}{\nonumber}

\newcommand{\sqrtg}{\sqrt{\rule{0pt}{0.8em}-g}}
\renewcommand{\L}{\Lambda}
\renewcommand{\S}{\Sigma}
\newcommand{\E}{{\cal E}}

\newcommand{\der}[2]{\frac{\partial{#1}}{\partial{#2}}}

\newcommand{\indU}[1]{{\mbox{\raisebox{0.1em}{\tiny$#1$}}}}
\newcommand{\indL}[1]{{\mbox{\raisebox{-0.15em}{\tiny$#1$}}}}

\newcommand{\palka}{\rule[-1em]{0.4pt}{2.3em}_{\,0}}

\begin{document}

%\maketitle
%%%%%%%%%%%%%%%%%%%%%%%%%%%%%%%%%%%%%%%%%%%%%%%%%%%%%%%%%%%%%%%%%%%%%%
\begin{titlepage}

    \thispagestyle{empty}
    \begin{flushright}
        \hfill{CERN-PH-TH/2009-080}\\
        \hfill\phantom{UCB-PTH-08/76}\\
        \hfill\phantom{SU-ITP-08/29}\\
    \end{flushright}

\vspace{3em}
\begin{center}
{\LARGE{\textbf{
Black hole entropy, flat directions \\[0.5em] and higher derivatives
}}}\\

\vspace{3em}

{{\textbf{S. Bellucci${}^1$, S. Ferrara${}^{1,2,3}$, A. Shcherbakov${}^{4,1}$, A. Yeranyan${}^{1,5}$}}}\\

\vspace{3em}
{${}^1$ \it INFN - Laboratori Nazionali di Frascati, \\
Via Enrico Fermi 40,00044 Frascati, Italy\\
\texttt{bellucci, ashcherb, ayeran@lnf.infn.it}}

\vspace{1em}
{${}^2$ \it Physics Department, Theory Unit, CERN, \\
CH 1211, Geneva 23, Switzerland\\
\texttt{sergio.ferrara@cern.ch}}

\vspace{1em}
{${}^3$ \it Department of Physics and Astronomy,\\
 University of California, Los Angeles, CA USA}

\vspace{1em}
{${}^4$ \it Museo Storico della Fisica e\\
        Centro Studi e Ricerche ``Enrico Fermi"\\
        Via Panisperna 89A, 00184 Roma, Italy}

\vspace{1em}
{${}^5$ \it Department of Physics, Yerevan State University,\\
Alex Manoogian St., 1, Yerevan, 0025, Armenia}

\end{center}
\vspace{2em}
\begin{abstract}
\vspace{1em}
Higher order derivative corrections to the Einstein--Maxwell
action are considered and an explicit form is found for the
corrections to the entropy of extremal black holes. We speculate
on the properties of these corrections from the point of view of
small black holes and in the case when the classical black hole
potential exhibits flat directions.
%
%In the latter case the
%corrections to the entropy contain values of the moduli at
%infinity which are not determined by the the attractor mechanism
%equations. This means that the attractor mechanism paradigm might
%fail in the presence of higher derivative corrections.
%Nevertheless, an accurate analysis shows that the attractor
%mechanism remains valid in the presence of corrections owing to a
%certain feedback when classically undefined moduli get fixed by
%quantum corrections.
%
%
%The relationship between two methods to
%calculate the black hole entropy~-- the so-called black hole
%potential approach and entropy function formalism -- is studied.
%Namely, it is shown here how the black hole potential is recovered
%from the entropy function and how the stability of the attractor
%mechanism equations can be studied entirely within the entropy
%function formalism. We consider in details an example of a dyonic
%black hole in heterotic string theory compactified on~$T^6$
%or~$K3\times T^2$: there we find two new~$\alpha{\,}'$-exact solutions
%and study their stability. Also we address the issue of the
%stability of the previously known solutions which turn out to be
%stable. A particular attention is paid to the stability of small
%black holes and it is found that in the considered example the
%Hessian matrix possesses negative eigenvalues. Hence, the
%corresponding solutions are unstable.
%
A particular attention is paid to the issue of stability of several solutions,
including large and small black holes by using properties of the Hessian matrix
of the effective black hole potential. This is done by using a model independent
expression for such matrix derived within the entropy function formalism.

\end{abstract}

\end{titlepage}
%%%%%%%%%%%%%%%%%%%%%%%%%%%%%%%%%%%%%%%%%%%%%%%%%%%%%%%%%%%%%%%%%%%%%%

%\tableofcontents
\section{Introduction}
Black hole physics provides us with a vast variety of phenomena
for testing underlying ideas of theoretical physics. This explains
the constant attention to this topical issue, as well as the burst
of interest any time a new concept emerges in this area of
research. Superstring/M-theory has enriched physics with new ideas
and currently is the main subject of interest in mathematical
physics. It allowed to intertwine different areas of theoretical
physics, e.g. gravity and conformal field theories.

There is a well known correspondence between black hole mechanics
and thermodynamics~\cite{Bekenstein:1973ur} that relates
geometrical characteristics of a black hole to thermodynamical
ones. One of them, the subject of our interest here,~-- entropy~--
can be calculated using the Wald formula~\cite{Wald:1993nt}. To be
able to interpret this quantity as a genuine entropy, it should be
confirmed by statistical physics calculations. This was
accomplished by performing counting of the
microstates~\cite{Strominger:1996sh,Strominger:1997eq} in the
``classical'' limit.  At this point, there arises a question of
comparison between microstate and ``macrostate'' entropies beyond
the classical approximation.

For extremal black holes the method of calculation of macrostate
entropy was proposed in~\cite{Ferrara:1997tw}. It relies on the
classical Einstein-Maxwell action and reduces the problem to
finding a black hole potential which encodes all the information
needed to obtain the entropy, which is given by the value of the
black hole potential at its critical point. This method  allows
one to trace back the evolution of the scalar fields in the whole
space, but might become difficult when considering higher order
corrections which originate from the string coupling and the
~$\alpha{\,}'$ expansion of the superstring action.

For this purpose another approach was proposed
in~\cite{Sen:2005wa}. It deals with horizon values of the fields
present in the theory by means of the so-called entropy function.

Taking into account terms coming from the superstring action (such
as the Chern--Simons term) allows one to achieve a matching
between black holes microstate and ``macrostate'' entropies
\cite{Prester:2008iu}.

Higher order corrections might be of interest from the classical
point of view due to the following problems:
\begin{enumerate}
\item small black hole~-- black hole with vanishing classical
entropy and non-vanishing values of mass.
Once classically one obtains a zero value of the entropy, one
naturally poses the question what happens when taking into
consideration higher order corrections as well.
\item flat directions of the black hole potential and stability of the critical points.
The presence of flat directions reflects the symmetry of the
scalar manifold. Therefore it is interesting to know how the
symmetry gets modified in the presence of higher order corrections
and, as a consequence, how flat directions get distorted and
whether the resulting critical point is stable or not.
\end{enumerate}
The problem of small black holes is studied from different points of view~\cite{Sen:2005kj}.

In order to investigate the influence of higher order corrections
on the solutions to the attractor mechanism equations~\cite{Ferrara:1995ih}, we make use
of the black hole potential  approach and the entropy function
formalism. We established a direct relation between these methods
and found how the black hole potential is related to the entropy
function. In addition, we derive a formula~(\ref{Hess}) for the Hessian
matrix of the black hole potential completely within the entropy
function formalism. A distinctive feature of this formula is that
it can be easily used even if deriving the expression of the black
hole potential from the entropy function proves to be problematic.

Previously, in the entropy function formalism not much attention
was paid to the stability issues, due to the fact that a
truncation over the ``problematic'' sectors (those where the flat
directions reside, e.g. the axionic one in the~$stu$ model) was
always considered. Having at disposal the formula for the Hessian
matrix allows us to study the stability of the complete, non
truncated version of the theory.

In the present work, we consider the most general case of higher
order derivative corrections to the Einstein-Maxwell action
involving both Riemann curvature tensor~$R_{\mu\nu\rho\sigma}$ and
electromagnetic field strength tensor~$F^\indU{\Lambda}_{\mu\nu}$
in any Lorentz covariant combinations. From dimensional analysis
it follows that the~$n^{\mbox{\scriptsize th}}$ order correction should have the
form~$(R_{\mu\nu\rho\sigma})^{m}
(F^\indU{\L}_{\mu\nu})^{2(n-m+1)}$ with~$m=0,\ldots,n+1$. The
indices are supposed to be contracted in all possible ways by
means of a metric~$g_{\mu\nu}$. The number of such terms grows
very fast with respect to~$n$, but in the entropy function
formalism they all lead to a much smaller number of independent
combinations. In this way we succeed in finding the corrections to
the black hole entropy induced by the higher order derivative
corrections to action.

As it is known, usually, in order to calculate up
to~$n^{\mbox{\scriptsize th}}$ order the value of a function at a
critical point, one should know the  solution to the criticality
condition up to~$(n-1)^{\mbox{\scriptsize th}}$ order. Examining
the form of the corrections to the entropy, one might easily
observe that there appears a problem starting from the first
order. Namely, the first order correction to the entropy is
supposed to be determined by the classical solution to the
attractor mechanism equations. Once the classical black hole
potential possesses flat directions, not all moduli are determined
by the attractor mechanism equations, though this ambiguity does
not affect the classical value of the entropy. But in the first
order this ambiguity shows up: the entropy becomes dependent on
the undetermined moduli. Evidently this fact contradicts the
second law of black hole thermodynamics and the attractor
mechanism paradigm. Nevertheless, a thorough analysis of the
perturbative corrections shows that there exists a sort of
feedback from the first order solutions on the classical ones. In
general, in order to calculate the entropy up
to~$n^{\mbox{\scriptsize th}}$ order, one should have the solution
up to~$n^{\mbox{\scriptsize th}}$ order. As it was mentioned
above,~$n^{\mbox{\scriptsize th}}$ order solution does not show up
in the expression for the entropy (at least
to~$n^{\mbox{\scriptsize th}}$ order), but it might not exist in
the presence of flat directions. This is related to the fact that,
in order to find the~$n^{\mbox{\scriptsize th}}$ order solution,
one has to solve a system of linear non-homogeneous equations with
vanishing determinant. This system contains as parameters moduli
which are not defined in the previous orders. Some of these
parameters get fixed when requiring the system of equations to be
consistent. In this way the equations to~$n^{\mbox{\scriptsize
th}}$ order fix the solutions in previous orders. Exactly for this
reason the attractor mechanism remains valid in the presence of
higher order derivative corrections, as well.

We considered in detail an example of a~$stu$ dyonic black hole in
heterotic string theory compactified on~$T^6$ or~$K3\times T^2$.
In addition to previously known solutions~\cite{Prester:2008iu,Sahoo:2006pm}, we found two new
solutions. They are both non-BPS and one of them, moreover,
corresponds to a state with vanishing central charge~$Z$~\cite{Bellucci:2006xz}. In the
classical limit the latter solution turns out to be stable.
Therefore we dwell mainly on the stability of the non-BPS~$Z\neq
0$ solution, which classically has two vanishing and four positive
eigenvalues. We find that, when the corrections are turned on, one
of the previously vanishing eigenvalues remains zero, while the
other becomes positive.

We studied as well the small black hole limit and found that the
two new solutions mentioned above do not allow for such a limit,
while the previously known ones do.  Therefore we studied the
stability of these solutions, as well, and found that for both of
them in the small black hole limit the Hessian matrix has always
at least one negative eigenvalue.

The paper is organized as follows. In section~\ref{basics} we
establish a relation between the two methods for calculating the
black hole entropy ~(\ref{VbhEntr}). The main result of this
section is the formula~(\ref{Hess}) yielding the Hessian matrix
expressed in terms of the entropy function. In section~\ref{hoc}
we introduce the higher order corrections to the action, discuss
their explicit form and derive the expressions for the black hole
effective potential~(\ref{Veff}) and entropy~(\ref{VPQ}), up to
the second order. We speculate on properties of the corrections to
the entropy both in the case of small black holes and when the
classical black hole potential has flat directions. In
section~\ref{example} we revisit a well known example of heterotic
string theory compactified on~$T^6$ or~$K3\times T^2$ without
neglecting axions. We demonstrate here that the corrections to the
entropy~(\ref{Entropy2}) do not depend on values of the scalars at
infinity. We derive two new non-BPS solutions and investigate
their stability, as well as the stability of the previously known
ones. A special attention is paid to small black holes and the
issue of their stability.

\section{Basics}\label{basics}

Since the paper~\cite{Ferrara:1997tw}, it has been known how to
calculate the entropy of a wide class of extremal black holes. There
an Einstein--Maxwell model coupled to scalar fields was considered
\be\label{EMaction}
S=\int d^4 x \sqrtg \left[ -\frac{R}2  + \frac12 G_{ab}(\phi)\,  g^{\mu\nu} \, \partial_{\mu} \phi^a \partial_{\nu}
\phi^b
    -\frac14 \mu_\indL{\L\S}(\phi) F^\indU{\L}_{\mu\nu} F^\indU{\S}{}^{\mu\nu}
    -\frac14 \nu_\indL{\L\S}(\phi) F^\indU{\L}_{\mu\nu} *F^\indU{\S}{}^{\mu\nu}\right]
\ee and it was shown that the entropy of a black hole in a static,
spherically symmetric and asymptotically flat space-time is
completely determined by the so-called~\emph{black hole potential},
which for the model~(\ref{EMaction}) has the form
\be\label{VBH}
V_{BH}(\phi) =  -\frac12(p^\indU{\Lambda},\,q_\indL{\Lambda})\left(
\begin{array}{cc}
\mu_\indL{\Lambda \Sigma}+\nu_\indL{\Lambda \Gamma
}\mu^{\indU{\Gamma\Pi}}\nu_\indL{\Pi\Sigma } &
    \nu_\indL{\Lambda \Gamma }\mu^\indU{\Gamma \Sigma } \\[0.5em]
\mu^\indU{\Lambda \Gamma }\nu_\indL{\Gamma \Sigma } &
    \mu^\indU{\Lambda\Sigma }
\end{array}
\right) \left(
\begin{array}{c}
p^\indU{\Sigma } \\
q_\indL{\Sigma }
\end{array}
\right). \ee
The entropy within this framework is given by the
value of the black hole potential at the horizon
\be\label{entrVal} S =\pi V_{BH}(\phi_h), \ee where the moduli
take critical values obtained from \be\label{critVBH}
\frac{\partial V_{BH}}{\partial
\phi^a}\,\rule[-0.75em]{0.5pt}{2.2em}_{\,\phi=\phi_h} = 0. \ee
This approach naturally allows us to examine the stability (which
guarantees that the expression in~(\ref{entrVal}) refers to the
entropy of a physical object) of the critical points by checking
the positive definiteness of the Hessian matrix
$$H_{ab} = \frac{\partial^2 V_{BH}}{\partial \phi^a \partial
\phi^b}\,\rule[-0.75em]{0.5pt}{2.2em}_{\,\phi=\phi_{h}},$$
and it
gives us the possibility to establish the presence of flat
directions of the black hole potential. Classically, the flat
directions are not fraught with any problem. Namely, the
criticality condition of~$V_{BH}$ might not fix uniquely the
values of the scalar fields, and some combinations of the fields
might remain free, while the value of the entropy does not depend
on these combinations. The problem might be hidden in quantum
correction terms which, in general, might destabilize the flat
direction (transforming it into a saddle critical point).

The presence of the additional terms~-- whether of quantum or
classical origin~-- in the action~(\ref{EMaction}) leads to a
modification of the black hole potential; there arises the
so-called effective black hole~$V_{eff}$ potential, whose extreme
value is equal to the entropy of the black hole and whose critical
points give the values of the scalar fields at the black hole
horizon. For the case of model~(\ref{EMaction}) the effective
potential~$V_{eff}$ is equal to the black hole one~$V_{BH}.$

An alternative approach~-- the entropy function formalism~-- for
calculating the entropy of a black hole with higher order
derivative corrections was proposed in~\cite{Sen:2005wa,
Sen:2005iz}. It may also be applied to calculate the entropy of a
large class of systems with additional degrees of freedom, e.g.
rotating black holes, black strings, multi-center black hole
configurations. Since we are also interested in stability issues,
we first relate this approach to the black hole potential one and
derive the concept of the Hessian matrix in the entropy function
formalism. Let us recall that the latter is based on putting the
theory on the near-horizon background geometry and constructing
the so-called \emph{entropy function} which can be easily written
down once the Lagrangian is known
$$
\ds \E(E^I,\phi^a) = 2\pi \left[ e^\indU{\L} q_\indL{\L} - \int d
\theta d \varphi \sqrtg {\cal L}\right],
$$
where~$E^I$ stands for all fields but the moduli. In this
formalism the black hole entropy is given by the value of the
entropy function at the horizon. The horizon values~$E^I_h$ and~$\phi^a_h$ of all fields
are determined by resolving all criticality conditions
simultaneously \be\label{eFunExtr} \frac{\partial \E}{\partial
E^I}\,\rule[-0.9em]{0.5pt}{2.4em}_{\,\scriptsize\ba{l} E=E_{h}\\
\phi=\phi_{h} \ea} = 0 ,\quad\mbox{(a)}\qquad \qquad \frac{\partial
\E}{\partial
\phi^a}\,\rule[-0.9em]{0.5pt}{2.4em}_{\,\scriptsize\ba{l} E=E_{h}\\
\phi=\phi_{h} \ea} =0. \quad\mbox{(b)} \ee

Since the entropy function formalism is more general, it
encompasses the black hole potential approach. One can obtain the
black hole potential from the entropy function as follows:
\be\label{VbhEntr} V_{BH}(\phi) = \pi^{-1}\, \E(E^I(\phi),\phi)
\ee where~$E^I(\phi)$ denotes the solution to the
equations~$\partial \E /\partial E^I = 0$ in terms of the moduli.
Of course, one can find the black hole potential, provided it is
possible to resolve this equation explicitly. For example, in
order to reproduce the black hole potential~(\ref{VBH}), one
should fix the near horizon geometry as~$AdS_2 \times S^2$
\be\label{AdS2S2} ds^2 =- v_1 \left( r^2 dt^2 - \frac{dr^2}{r^2}
\right) + v_2 \left( d\theta^2 + \sin^2 \theta d \varphi^2
\right), \ee resolve Bianchi identities for the electromagnetic
fields \be\label{EM} F^\indU{\L}_{01} = - e^\indU{\L}(r), \qquad
F^\indU{\L}_{23} = p^\indU{\L} \sin\theta \ee
and substitute
the explicit dependence of~$E^I=(e^\indU{\L},v_1,v_2)$ on~$\phi^a$
\be\label{sol1}
\ds e^\indU{\L} = -\mu^\indU{\L\S} \left( q_\indL{\S} - \nu_\indL{\S\Omega} p^\indU{\Omega}\right),\quad%\\[0.6em]
\ds v_1 = v_2 = - \frac12\,\left[\rule{0pt}{1em} \mu_\indL{\L\S} \,
p^\indU{\L} p^\indU{\S}
    + \mu^\indU{\L\S}
    \left( q_\indL{\L} - \nu_\indL{\L\L'}\, p^\indU{\L'}\right)
    \left( q_\indL{\S} - \nu_\indL{\S\S'}\, p^\indU{\S'} \right) \right]
\ee back into the entropy function.

In many cases one cannot resolve the criticality conditions~(\ref{eFunExtr}a) and hence derive the black hole potential, therefore the entropy function formalism is
more general. In order to make it self-contained one should be
able to answer the question about the stability, that is to
calculate the Hessian matrix in terms of the entropy function
only. It becomes important when one cannot deduce the explicit
dependence of~$E^I$ on~$\phi^a$.

From the formula~(\ref{VbhEntr}) it follows that the Hessian matrix
is equal to
$$H_{ab} = \frac1\pi\,\left[ \frac{\partial^2 \E}{\partial E^I \partial E^J} \frac{\partial E^I}{\partial \phi^a}  \frac{\partial E^J}{\partial \phi^b}
    + \frac{\partial^2 \E}{\partial E^I \partial \phi^a} \frac{\partial E^I}{\partial \phi^b}
    + \frac{\partial^2 \E}{\partial E^I \partial \phi^b} \frac{\partial E^I}{\partial \phi^a}
    + \frac{\partial^2 \E}{\partial \phi^a \partial \phi^b} \right]\rule[-1.7em]{0.pt}{2.4em}_{\scriptsize\ba{l} E=E_{h}\\ \phi=\phi_{h} \ea}.
$$
Supposing that one does not have the explicit form of~$E^I(\phi)$,
let us try to express~${\partial E^I}/{\partial \phi^a}$ through
the entropy function. For this purpose it is sufficient to
differentiate~$\partial \E /\partial E^I = 0$ with respect
to~$\phi^a$, considering~$E^I$ as a function of~$\phi^a$:
$$
\frac{\partial^2 \E}{\partial E^I \partial E^J} \frac{\partial
E^J}{\partial \phi^a} + \frac{\partial^2 \E}{\partial E^I \partial
\phi^a} = 0 \quad \Rightarrow \quad \frac{\partial E^I}{\partial
\phi^a}
    = - \left( \frac{\partial^2 \E}{\partial E^I \partial E^J} \right)^{-1} \frac{\partial^2 \E}{\partial E^J \partial \phi^a}
$$
which yields the following expression for the Hessian matrix:
\be\label{Hess}
\ds H_{ab} =\frac1\pi\,\left[
     \frac{\partial^2 \E}{ \partial \phi^a \partial \phi^b }
     - \left(\frac{\partial^2 \E}{\partial E^I \partial E^J}\right)^{-1}
      \frac{\partial^2 \E}{ \partial E^I \partial \phi^a} \frac{\partial^2 \E}{ \partial E^J \partial \phi^b}
     \right]\rule[-1.7em]{0.pt}{2.4em}_{\scriptsize\ba{l} E=E_{h}\\ \phi=\phi_{h} \ea}.
\ee Notice, that this expression is quite simple to deal with. It
requires the knowledge of the derivatives of the entropy function
which are easily calculable. Amusingly enough, for studying the
stability of the solutions, one does not have to be able to
resolve equations~$\partial \E /\partial E^I = 0$ in terms of the
moduli in order to reconstruct the black hole potential. This
issue becomes important when considering higher order derivative
corrections.

\section{Higher order corrections}\label{hoc}
Now let us expand the considerations we made above by adding
higher order derivative terms to the action~(\ref{EMaction}):
$$
\ba{l}
\ds S=\int d^4 x \sqrtg \left[ -\frac{R}2 + \frac12
G_{ab}(\phi)\,  g^{\mu\nu} \, \partial_{\mu} \phi^a \partial_{\nu}
\phi^b
    -\frac14 \mu_\indL{\L\S}(\phi) F^\indU{\L}_{\mu\nu} F^\indU{\S}{}^{\mu\nu}\right.\\[0.6em]
\ds \phantom{S=\int d^4 x \sqrtg \left[ -\frac{R}2 \right] }\left.
\ds    -\frac14 \nu_\indL{\L\S}(\phi) F^\indU{\L}_{\mu\nu} *F^\indU{\S}{}^{\mu\nu}\right]
    + \int d^4 x \sqrtg \, {\cal L}_H(R_{\mu\nu\lambda\sigma},g^{\mu\nu},F^\indU{\L}_{\mu\nu},\phi)
\ea\eqno{(\ref{EMaction}')}
$$
The last term corresponds to higher order derivative corrections
coming, for example, from the~$\alpha{\,}'$-expansion of the heterotic
string action~\cite{Metsaev:1987zx}; its form will be specified
later.

The entropy function corresponding to the model~(\ref{EMaction}${}^\prime$) acquires an additional term
\be\label{eFun2}
\ba{l}
\ds \E(e,v,\phi) = %2\pi \left[ e^\indU{\L} q_\indL{\L} - \int d \theta d \phi \sqrtg {\cal L}\right]=\\[0.7em]
%\ds \phantom{\E(e,v,z,\bar z) } =
2\pi \left[\,e^\indU{\L} q_\indL{\L} +  v_2 - v_1
    -\frac12\, \mu_\indL{\L\S} \left( \frac{v_1}{v_2} p^\indU{\L} p^\indU{\S}
    - \frac{v_2}{v_1}\,  e^\indU{\L} e^\indU{\S} \right)
    - \nu_\indL{\L\S}\, e^\indU{\L}\, p^\indU{\S} \right] + \E_H
\ea \ee which is defined as an integral calculated on~$AdS_2
\times S^2$ background~(\ref{AdS2S2}) with resolved Bianchi
identities~(\ref{EM})
\be\label{fH}
\E_H(v_1,v_2,p,e,\phi) = -\frac14 \int_{S^2} d\theta d \varphi \sqrtg\, {\cal L}_H.%= v_1 v_2 \; {\cal L}_H(v_1,v_2,p,e,z,\bar z).
\ee Since the contribution from the term~$\E_H$ is supposed to
come from the perturbative expansion of the superstring action, we
present it in the following form: \be\label{f1f2} \E_H=\alpha{\,}'
\E_1 + \alpha{\,}'{}^2 \E_2 + O(\alpha{\,}'{}^3). \ee Our goal is to
construct the black hole potential corresponding to the
action~(\ref{EMaction}$'$). In what follows we reserve the term
``black hole potential'' for a potential corresponding to the
classical action~(\ref{EMaction}), and the term ``effective black
hole potential''~-- to the action~(\ref{EMaction}$'$) with higher
order derivative corrections.

As it was demonstrated in the previous section, to obtain the
effective potential one should eliminate in the entropy function
all fields except the moduli, using their equations of motion. For
this purpose we represent the entropy function in the form
\be
\E(e,v,\phi) = \E_0(e,v,\phi) + \alpha{\,}' \E_1(e,v,\phi) +
\alpha{\,}'{}^2 \E_2(e,v,\phi) + O(\alpha{\,}'{}^3)
\ee
and for simplicity
introduce again a notation~$E^I=(v_1,v_2,e^\indU{\L})$ for the
``superfluous'' fields. It can be easily shown that the effective
potential, up to the second  order in~$\alpha{\,}'$, is then given by
\be  V_{eff}(\phi) = \frac1{\pi}\left[ \E(E^I,\phi^a)
    - \frac{\alpha{\,}'{}^2}2\, \left( H_{IJ}\right)^{-1} \der{\E_1}{E^\indU{I}} \der{\E_1}{E^\indU{J}} \right]_{\,E=E_0} + O(\alpha{\,}'{}^3), \quad
    H_{IJ} = \frac{\partial^2 \E_0}{\partial E^\indU{I} \partial E^\indU{J}}
\ee where~$E_0$ stands for the ``classical''
solutions~(\ref{sol1}) for the fields~$v_{1,2}$ and~$e^\indU{\L}$
in terms of the moduli. The matrix~$H^\indU{-1}$ has a following
block form: \be H^\indU{-1} =-\frac1{2\pi}\, \left( \ba{ccc}
2 V_{BH}  & V_{BH}  & e^\indU{\L}\\
V_{BH}    & 0     & e^\indU{\L}\\
e^\indU{\S} &  e^\indU{\S} & -\mu^\indU{\L\S} \ea \right)
\ee
with~$V_{BH}$ and~$e^\indU{\L}$ given in eqs.~(\ref{VBH}) and~(\ref{sol1}).
Therefore we may represent the effective potential in the form
\be\label{Veff}
\ba{l}
\ds V_{eff}(\phi) = V_{BH}(\phi) + \frac{\alpha{\,}'}{\pi}\, \E_1(E_0(\phi),\phi) + \frac{\alpha{\,}'{}^2}{\pi}\, \E_2(E_0(\phi),\phi)  \\[0.7em]
\ds \phantom{V_{eff}(\phi) =}
    -\frac{\alpha{\,}'{}^2}{4\pi^2} \left[
    \mu^\indU{\L\S}\,\der{\E_1}{e^\indU{\L}}\der{\E_1}{e^\indU{\S}}
    - \frac2{V_{BH}} \left(v_1 \der{\E_1}{v_1} + e^\indU{\L}\der{\E_1}{e^\indU{\L}} \right)\left(v_1 \der{\E_1}{v_1} + v_2 \der{\E_1}{v_2} \right)
    \right] + O(\alpha{\,}'{}^3)
\ea
\ee
with the right hand side, obviously, being calculated on the classical solution~(\ref{sol1}).

\subsection{Explicit form of the corrections}
Now we try to make our considerations more specific by defining
the form of the corrections. Although we are interested only up
to~$\alpha{\,}'{}^2$ order corrections to the effective potential, the
formulae of this section might be easily written down for the
corrections of any order.

To the~$\alpha{\,}'{}^1$ order, the possible corrections allowed by
dimensinal analysis have the following schematic structure:
$${\cal L}^{(1)}_H \quad\sim\quad (R_{\mu\nu\rho\sigma})^2 \quad + \quad R_{\mu\nu\rho\sigma} (F_{\mu\nu})^2 \quad + \quad (F_{\mu\nu})^4,$$
where the Lorentz indices are supposed to be contracted with a
proper number of inverse metric~$g^{\mu\nu}$ in all possible ways.
This expression induces the following form of the first order
correction to the entropy function: \be \E_1 = \E_{R^2} +
\E_{RF^2} + \E_{F^4}. \ee Calculated on the~$AdS_2 \times S^2$
background~(\ref{AdS2S2}) and with the field strength given by
eqs.~(\ref{EM}), the terms composing~$\E_1$ are then given by
\be\label{alpha1} \ba{l}
\ds \E_{R^2\phantom{F}}  \sim  v_1 v_2 \left[ \frac{\alpha^{(1)}}{v_1^2} + \frac{\alpha^{(2)}}{v_1 v_2} + \frac{\alpha^{(3)}}{v_2^2}\right]\,,\\[1em]
\ds \E_{RF^2} \sim v_1 v_2 \left[
    \left( \frac{\alpha^{(4)}_\indL{\L\S}}{v_1} + \frac{\alpha^{(5)}_\indL{\L\S}}{v_2} \right)
        \frac{e^\indU{\L} e^\indU{\S}}{v_1^2}
    + \left( \frac{\alpha^{(6)}_\indL{\L\S}}{v_1} + \frac{\alpha^{(7)}_\indL{\L\S}}{v_2} \right)
        \frac{e^\indU{\L} p^\indU{\S}}{v_1 v_2}
    + \left( \frac{\alpha^{(8)}_\indL{\L\S}}{v_1} + \frac{\alpha^{(9)}_\indL{\L\S}}{v_2} \right)
        \frac{p^\indU{\L} p^\indU{\S}}{v_2^2}\right],\\[1em]
\ds \E_{F^4\phantom{R}} \sim v_1 v_2 \left[
    \alpha^{(10)}_\indL{\L\S\Pi\Omega} \frac{e^\indU{\L} e^\indU{\S} e^\indU{\Pi} e^\indU{\Omega}}{v_1^4}
    + \alpha^{(11)}_\indL{\L\S\Pi\Omega} \frac{e^\indU{\L} e^\indU{\S} e^\indU{\Pi} p^\indU{\Omega}}{v_1^3 v_2}
    + \dots
    + \alpha^{(14)}_\indL{\L\S\Pi\Omega}  \frac{p^\indU{\L} p^\indU{\S} p^\indU{\Pi} p^\indU{\Omega}}{v_2^4}\right]\,,
\ea
\ee
where all~$\alpha^{(n)}$ are functions of~$\phi^a$.

Very schematically we present, as well, the second order
corrections to the entropy function
\be\label{alpha2} \ba{l} \ds
\E_{R^3} \sim  v_1 v_2 \left[ \frac1{v_1^3} + \frac1{v_1^2 v_2} +
\frac1{v_1 v_2^2}+ \frac1{v_2^3}\right]\,, \quad \ds \E_{R F^4}
\sim v_1 v_2 \left[ \frac1{v_1} + \frac1{v_2} \right]
    \left[ \frac{e^4}{v_1^4} + \frac{e^3 p}{v_1^3 v_2}
    + \dots + \frac{p^4}{v_2^4}\right],\\[1em]
\ds \E_{R^2 F^2} \sim v_1 v_2 \left[ \frac1{v_1^2} + \frac1{v_1 v_2} + \frac1{v_2^2} \right]
    \left[ \frac{e^2}{v_1^2} + \frac{e p}{v_1 v_2} + \frac{p^2}{v_2^2}\right], \quad
\ds \E_{F^6} \sim v_1 v_2 \left[\frac{e^6}{v_1^6}
    +  \frac{e^5 p}{v_1^5 v_2}
    + \dots + \frac{p^6}{v_2^6}\right]\,.
\ea \ee Here the coefficients are assumed to depend on the scalar
fields~$\phi^a$, but their dependence is not written explicitly
for the sake of simplicity. The function~$\E_2$~(\ref{f1f2}), in
turn, is given by the sum
$$\E_2 = \E_{R^3}+\E_{R^2F^2} + \E_{RF^4} + \E_{F^6}. $$
Each of the functions of eqs.~(\ref{alpha1}) and~(\ref{alpha2})
turns out to be an eigenfunction of the operator appearing in the
expression for the effective potential~(\ref{Veff})
$$ \left[ v_1\frac{\partial}{\partial v_1} + v_2\frac{\partial}{\partial v_2} \right] \E_{R^n F^m} = \left( 2-n-m \right) \E_{R^n F^m}.$$
This allows us to rewrite the effective potential~(\ref{Veff}) in
the following form:
$$\pi V_{eff}(\phi) = \E(e,v,\phi) - \frac{\alpha{\,}'{}^2}{4\pi} \left[
    \mu^\indU{\L\S}\,\der{\E_1}{e^\indU{\L}}\der{\E_1}{e^\indU{\S}}
    + \frac2{V_{BH}}\,\left(\strut \E_{RF^2} + 2\,\E_{F^4} \right) \left(v_1 \der{\E_1}{v_1} + e^\indU{\L}\der{\E_1}{e^\indU{\L}} \right)
     \right],
\eqno{(\ref{Veff}')} $$
where, as before, the right hand side is supposed to be taken on the solution~(\ref{sol1}).

\subsection{Corrections to the value of the entropy}
In the previous section we derived the effective black hole potential which encodes the
 information about the entropy of a black hole. Now we are going to calculate the value of the
 entropy up to the second order in~$\alpha{\,}'$.
 So, we should extremize the effective black hole potential with respect to the scalar fields
\be\label{critVphi} \der{V_{eff}}{\phi^a} =  0. \ee In order to
resolve these equations perturbatively let us expand the effective
potential and the scalar fields in series on~$\alpha{\,}'$ up to the
second order \be\label{expansionsVphi} V_{eff} = V_{0} + \alpha{\,}'
V_1 + \alpha{\,}'{}^2 V_2 + O(\alpha{\,}'{}^3), \qquad \phi^a = \phi_0^a +
\alpha{\,}' \phi_1^a + \alpha{\,}'{}^2 \phi_2^a + O(\alpha{\,}'{}^3). \ee The
expansion of~$V_{eff}$ is nothing but a concise form of the
eq.~(\ref{Veff}$'$). Being written in full details it gives
\be\label{VvsE} \ba{l}
\ds V_0 = \pi^{-1}\, \E_0 = V_{BH}, \qquad \ds V_1 =  \pi^{-1}\, \E_1, \\[0.5em]
\ds V_2 = \frac1{\pi}\, \E_2 - \frac1{4\pi^2} \left[
    \mu^\indU{\L\S}\,\der{\E_1}{e^\indU{\L}}\der{\E_1}{e^\indU{\S}}
    + \frac2{V_{BH}}\,\left(\strut \E_{RF^2} + 2\,\E_{F^4} \right) \left(v_1 \der{\E_1}{v_1} + e^\indU{\L}\der{\E_1}{e^\indU{\L}} \right)
\right] \ea \ee where all~$\E$-terms are calculated on the
``classical'' solution~(\ref{sol1}). After substitution of the
expansions~(\ref{expansionsVphi}) into the criticality
condition~(\ref{critVphi}) one immediately derives
\be\label{phiEqs} \ba{ll}
\ds \der{V_0}{\phi^a}\,\palka=0, & \ds \quad(a)\\
\ds \frac{\partial^2 V_0}{\partial\phi^a \partial\phi^b}\,\palka\, \phi^b_1
    = - \der{V_1}{\phi^a}\,\palka,& \ds \quad (b)\\
\ds \frac{\partial^2 V_0}{\partial\phi^a \partial\phi^b}\,\palka\, \phi^b_2
    = - \der{V_2}{\phi^a}\,\palka
    -  \frac{\partial^2 V_1}{\partial\phi^a \partial\phi^b}\,\palka\, \phi^b_1
    - \frac12 \frac{\partial^3 V_0}{\partial\phi^a \partial\phi^b \partial\phi^c}\,\palka\, \phi^b_1 \phi^c_1,& \ds\quad (c)
\ea \ee where the subscript zero means that the corresponding
expression should be taken upon~$\phi^a=\phi^a_0$. Although, in
order to obtain an extreme value of a function up to~$\alpha{\,}'{}^n$
order, it is enough to know a solution to criticality condition up
to~$\alpha{\,}'{}^{(n-1)}$ order, we wrote down as well an equation~(\ref{phiEqs}c)
defining the second order solution~$\phi^a_2$. It relates to a
subtle effect that we illustrate on a simpler example.

Let us suppose that we are interested in calculating the value of
the entropy up to~$\alpha{\,}'{}^1$ order. In this case we need to know
only a ``classical'' solution~$\phi_0^a$ to the
eq.~(\ref{phiEqs}a), so that one might think that
eqs.~(\ref{phiEqs}b) are of no importance, since they define the
value of the first order solution~$\phi_1^a$. The subtlety comes
out when the ``classical'' black hole potential~$V_0$ has a flat
direction. In this case the matrix of the second derivatives
(which is nothing but a Hessian matrix) becomes degenerate and not
all of~$\phi_0^a$ might be determined from~(\ref{phiEqs}a). The
undetermined scalar fields appear then in the right hand side
of~(\ref{phiEqs}b). This fact might make the system of
eqs.~(\ref{phiEqs}b) to become inconsistent, since the Hessian matrix is
degenerate. Considering this possibility as quite unphysical, we
impose the condition that~(\ref{phiEqs}b) be consistent.  This
condition might fix some of the previously
undetermined~$\phi_0^a$. If the eqs.~(\ref{phiEqs}b) turns out to
be consistent identically, it means that the symmetry of the first
order corrections to the effective potential coincides with the
symmetry of the ``classical'' effective potential.

When one is interested in corrections up to~$\alpha{\,}'{}^2$, then
one has to check also that~(\ref{phiEqs}c) is consistent. The
extreme value of the effective potential (that is, the entropy in
fact) is then given by the expression \be\label{VeffValue}
V_{eff.extr} = V_0(\phi_0) + \alpha{\,}' V_1(\phi_0) +
\alpha{\,}'{}^2\left[
    V_2(\phi_0) - \frac12 \der{V_1}{\phi^a}\palka \phi^a_1
    \right].
\ee Let us analyse this expression. First of all, the black hole
potential~(\ref{VBH}) is a homogeneous function of degree two in
the charges~$p^\indU{\L}$ and~$q_\indL{\L}$. Then, the
``classical'' solution~$\phi_0$ is homogeneous of zero degree on
the charges. It immediately follows from the fact that~$V_{BH}$ is
homogeneous and~$\phi_0$ is a solution to a homogeneous
equation~(\ref{critVBH}). In order to write down the dependence of
the entropy on the charges, let us for simplicity combine them to
form a symplectic charge vector
$$P^\indU{\L} = \left(p^\indU{\L}, q_\indL{\L} \right). $$
Calculating the expressions~$\E_{1,2}$ on the ``classical''
solutions according to formulae~(\ref{alpha1}) and~(\ref{alpha2})
and taking into account eqs.~(\ref{VvsE}),~(\ref{phiEqs})
and~(\ref{VeffValue}), one gets \be\label{VPQ}
\newcommand{\VBH}{S_0}
\ba{l}
\ds %V_{eff.extr}
 S = \VBH + \alpha{\,}' \left[ A + \frac1{\VBH}\, A_\indL{\L\S} P^\indU{\L} P^\indU{\S}
    + \frac1{\VBH^{\,2}}\, A_\indL{\L\S\Pi\Omega} P^\indU{\L} P^\indU{\S} P^\indU{\Pi} P^\indU{\Omega}\right]
    + \frac{\alpha{\,}'{}^2}{\VBH} \left[B + \frac1{\VBH}\, B_\indL{\L\S} P^\indU{\L} P^\indU{\S} \right.\\
\ds\phantom{S=S_0+} \left.
\ds + \frac1{\VBH^{\,2}}\, B_\indL{\L\S\Pi\Omega} P^\indU{\L} P^\indU{\S} P^\indU{\Pi} P^\indU{\Omega}
    + \frac1{\VBH^{\,3}}\, B_\indL{{\L}_1 \ldots {\L}_6} P^\indU{\L_1}\ldots P^\indU{\L_6}\right]
    + O(\alpha{\,}'{}^3).
\ea \ee
The coefficients~$A, A_\indL{\L\S}, A_\indL{\L\S\Pi\Omega}$ and~$B, B_\indL{\L\S}, \ldots$ are
composed of the previously introduced functions~$\alpha^{(n)}$, $\mu_\indL{\L\S}$ and~$\nu_\indL{\L\S}$ and
in general depend on the ``classical'' values of the scalar fields
that are homogeneous of the zero degree on the charges. The
explicit form of these coefficients depends on the model one
considers and, if one considers higher dimensional theory, on the
possible compactifications to four dimensions. Nevertheless, the
structure of the contributions coming from the higher derivatives
remains the same.

The generalization of this formula for higher order derivative
corrections is straightforward. We see that really all the series
is built out of two quantities: the classical value of the entropy
(proportional to~$V_{BH}$), which is quadratic in the charges for
the extremal black holes in~$D=4$, and the classical values of the
scalar fields, which are homogeneous of degree zero.

Formula~(\ref{VPQ}) is an agreement with the result obtained for
the one-loop correction to the Bekenstein--Hawking
entropy~\cite{Maldacena} when only the~$R^2$ term is present in
the action~(\ref{EMaction}$'$).

As we have noticed before, the coefficients~$A$ and~$B$ (with any
number of indices) in~(\ref{VPQ}) depend on the classical values
of the scalar fields. If the black hole potential~$V_{BH}$ has a
flat direction, this means that not all of the scalar fields get
fixed. This is not a problem for the~$V_{BH}$, since its value
does not depend on the scalars which are not fixed. But the first
order correction might depend on these not fixed values. This
means that black hole entropy depends on the values of the
non-fixed scalars which can change continuously. This fact
threatens the attractor mechanism paradigm and can induce a
violation of the second law of black hole thermodynamics. Really,
such a violation does not occur and the attractor mechanism
paradigm is preserved. As it has been already
explained, once a flat direction is present, the system of
eqs.~(\ref{phiEqs}b) becomes degenerate and to possess a solution
its right hand side should satisfy some consistency condition,
which might fix some of the previously free scalar
fields~$\phi^a_0$. If not all of the scalars~$\phi^a_0$ get fixed,
then the group symmetry of the higher derivative corrections is a
subgroup of the symmetry of the black hole potential, hence the
higher order corrections will not depend on these non-fixed
scalars.

Looking at the eq.~(\ref{VPQ}) one sees that higher order
corrections become singular when considering small black holes.
Once~$V_{BH}$ is equal to zero, the perturbative expansion is not
valid anymore. And there arises a question whether quantum effects
might cure small black holes. This will the subject of a
forthcoming analysis, whereas for the time being we limit
ourselves to just some remarks. In realistic models the
singularity might be removed when along with~$S_0 =  0$ the
coefficients~$A$ and~$B$ tend to zero as~$\phi^a = \phi^a_0$ in a
way such that the corresponding fractions in~(\ref{VPQ}) remain
finite.

\section{Application to the~$stu$ model}\label{example}
In order to shed light on specific features of the effect of quantum
corrections in the case of small black holes or when flat directions
are present, we are going to revisit some well known examples where
both such features are present. In particular we consider the~$stu$
model with higher order corrections stemming from compactification
of the heterotic string theory to four
dimensions~\cite{Prester:2008iu, Sahoo:2006pm}. The classical
non-BPS solution for this model possesses two flat
directions~\cite{Ferrara:2007tu} in the axionic sector of its moduli
space. In addition, this example allows us to investigate small
black holes.

In order to derive the form of the higher order corrections to
this model we take the action of the heterotic string
model~\cite{Metsaev:1987zx}, with all~$\alpha{\,}'$ order terms
calculated, and truncate it to six dimensions. Being afterwards
reduced to four dimensions it corresponds to the~$stu$ model with
higher order derivative terms.

We perform our computations directly in six dimensions since
dimensional reduction of the~$\alpha{\,}'$ corrections to four
dimensions might be a topic of independent research. Moreover,
working in six dimensions makes it easier to take into
consideration the gravitational Chern--Simons
term~\cite{Prester:2008iu,Sahoo:2006pm}.

We will follow the strategy given in~\cite{Sahoo:2006pm} and, in
order to simplify the comparison of the results, we will mostly
follow the notations by~\cite{Sahoo:2006pm}. The difference of our
consideration from that of~\cite{Sahoo:2006pm} is that we preserve
all the scalar fields~-- axions and dilatons~-- present in the
theory, since the axion fields turn out to play an important role
when it comes to the question of stability.

The action~\cite{Metsaev:1987zx} we start with is the bosonic part
of the heterotic string theory action truncated to six dimensions.
It describes the coupling of a two-form field~$B_\indL{MN}$ and a
dilaton field~$\Phi$ to six-dimensional Einstein gravity
\be\label{MT} \ba{l} \ds S=\frac{1}{32 \pi}\int d^6 x
\sqrt{\strut-G}\, e^{-\Phi} \left[ R + \partial_\indL{M} \Phi
\partial^\indU{M} \Phi
    - \frac1{12} H_\indL{MNK} H^\indU{MNK}  +\frac{\alpha{\,}'}{8}\;
    \bar{R}_\indL{MNKL} \bar{R}^\indU{MNKL} + \alpha{\,}'^3 \Delta\mathcal{L}_{3} \right] + \ldots
\ea
\ee
where
\be
\bar{R}^\indU{M}{}_\indL{NPQ} = R^\indU{M}{}_\indL{NPQ} + \nabla_\indL{[P} H^\indU{M}{}_\indL{Q]N}
 - \frac12 H^\indU{M}{}_\indL{R[P} H^\indU{R}{}_\indL{Q]N} \;.
\ee Here the term~$\Delta\mathcal{L}_{3}$ is of the forth order in
the curvature~$\bar{R}_\indL{MNKL}$. The field
strength~$H_\indL{PQR}$ includes also a gravitational
Chern--Simons term \be \ba{l}
\ds H_\indL{PQR} = \partial_\indL{P} B_\indL{QR}+\partial_\indL{Q} B_\indL{RP}+\partial_\indL{R} B_\indL{PQ} - 3 \alpha{\,}' \left(\Omega_\indL{PQR}+\mathcal{A}_\indL{PQR}\right),\\[0.5em]
\ds \Omega_\indL{PQR} = \frac12\, \Gamma^\indU{N}_\indL{PM}\partial_\indL{Q} \Gamma^\indU{M}_\indL{NR}
    + \frac13\, \Gamma^\indU{N}_\indL{PM} \Gamma^\indU{M}_\indL{QK} \Gamma^\indU{K}_\indL{NR} + \mbox{antisym. on }P,Q,R,\\[0.6em]
\ds\mathcal{A}_\indL{PQR} = \frac14 \, \partial_\indL{P} \left( \Gamma^\indU{M}_\indL{QN} H^\indU{N}{}_\indL{MR} \right)
    + \frac18\, H^\indU{M}{}_\indL{PN} \nabla_Q H^\indU{N}{}_\indL{MR}
    - \frac14\, R^\indU{MN}{}_\indL{PQ} H^\indU{RMN} \nn \\
\ds\phantom{\mathcal{A}_\indL{PQR}=}
    + \frac1{24}\, H^\indU{M}{}_\indL{PN} H^\indU{S}{}_\indL{QM} H^\indU{N}{}_\indL{RS} + \mbox{ antisym. on }P,Q,R.
\ea \ee As prescribed in~\cite{Sahoo:2006pm}, one may dualize the
field strength~$H_\indL{PQR}$ into a new one~$K_\indL{PQR}$ which
is an exact three-form
$$ K_\indL{PQR} = \partial_\indL{P} C_\indL{QR}+\partial_\indL{Q} C_\indL{RP}+\partial_\indL{R} C_\indL{PQ},$$
\noindent obeying, obviously, the Bianchi identities. To this end,
one adds the term
$$ \int d^6 x \varepsilon^\indU{MNKPQR} K_\indL{MNK}\left(\strut H_\indL{PQR} +3 \alpha{\,}' \Omega_\indL{PQR}\right) $$
into the original action~(\ref{MT}) and eliminates the field
strength~$H_\indL{PQR}$. We postpone the elimination
of~$H_\indL{PQR}$ for a while, for a reason to be clarified later.

Dealing with the Chern-Simons term is a little bit
tricky~\cite{Sahoo:2006pm} and consists in maintaining the Lorentz
covariance of the Lagrangian. We succeeded in singling out a
manifestly covariant part of the Lagrangian by adding total
derivative terms, in such a way that the second derivatives of the
four dimensional electromagnetic potentials originating in the six
dimensional metric disappear.

In order to be able to interpret the model~(\ref{MT}) as an
ancestor of the~$stu$ model, one should clarify the field content
of the theory. First of all, the six dimensional metric
tensor\footnote{we assume the splitting of the six dimensional
indices~$M,N,K,\ldots=0,1,\ldots,5$ into four
dimensional~$\mu,\nu,\rho,\ldots=0,1,2,3$ and two
dimensional~$m,n,k,\ldots=4,5$ ones}~$G_\indL{MN}$ when reduced to
four dimensions produces a four dimensional metric
tensor~$g_{\mu\nu}$, two vector fields~$A^{1,2}_\mu$ and three
scalar fields~-- two dilatons~$u_1, u_2$ and one axion~$c$. Then,
the two-form potential~$C_\indL{MN}$ produces a four dimensional
two-form~$C^{(4)}_{\mu\nu}$, two vector fields~$A^{3,4}_\mu$ and
one axion~$C_{mn} = b\, \epsilon_{mn}$. In four dimensions a
two-form~$C^{(4)}_{\mu\nu}$ is dual to a scalar; this duality
gives rise to another axion~$a$. Finally, the original scalar
field~$\Phi$ gives rise to a four dimensional dilaton~$u_s$. So,
we end up with a four dimensional metric tensor, four vector
fields and six real scalars~-- three axions~$a,b,c$ and three
dilatons~$u_1,u_2,u_s$ (more precisely, exponentials of the
dilatons).

The entropy function formalism in this case is based on putting the
theory on a six-dimensional background
\be
ds^2_{(6)} = ds^2_{(4)} + G_{mn} \left( dx^m + A_\mu^{m-3} dx^\mu\right) \left( dx^n + A_\nu^{n-3} dx^\nu \right),
\ee
where the
two-dimensional metric tensor~$G_{mn}$ is parameterized as
follows~\cite{Duff:1995sm}: \be
G_{mn}=  \left( \ba{cc} u_1^2 + c^2\, u_2^2 & -c\, u_2^2 \\
-c \,u_2^2 & u_2^2 \ea \right). \ee The connections~$A^{m-3}_\mu$
are chosen in the following form: \be A^1_\mu = \left( 2\, r\,
e^1, 0,0,0 \right), \qquad A^2_\mu = ( 2\, r\, e^2, 0,0, -
\frac{p^2}{2\pi}\cos\theta ). \ee Hence, in what follows they will
give the so-called magnetic configuration corresponding to
the~$D0-D4$ branes. The other two vector fields coming
from~$C_\indL{MN}$ are equal to \be A^3_\mu = \left(\strut\right.
\frac18\, r\, e^3, 0,0,0 \left.\strut\right), \qquad A^4_\mu =
\left(\strut\right. \frac18\, r\, e^4, 0,0, -
\frac{p^4}{32\pi}\cos\theta \left.\strut\right). \ee The electric
potentials~$e^\indL{\L}$ are dynamical fields; therefore we do not
put them equal to zero ad hoc, as their values are to be fixed
when minimizing the entropy function.

In the entropy function formalism one should fix the values of the
scalars at the horizon of the black hole. Thus we parameterize the
dilaton field as follows~\cite{Sahoo:2006pm}:
$$e^{-\Phi} = \frac{u_s}{64 \pi^2 u_1 u_2}   $$
From a four dimensional perspective the field
strength~$K_\indL{MNK}$ is related to the vector potentials and
axions in the following way:
$$
\ba{l}
\ds K_{\mu\nu\rho}= \frac{u_s}{64 \pi^2 u^2_1 u^2_2} \sqrtg g^{ \sigma \tau} \epsilon_{\tau \mu\nu\rho}\, a,_{\sigma}
    + 3 K_{[\mu\nu n} A^{n-3}_{\lambda]} + 3 K^{\phantom{n-3}}_{[\mu m n} A^{m-3}_\nu A^{n-3}_{\lambda]}\\
\ds K_{\mu\nu m} = - F^{m-1}_{\mu\nu} -  b\, \epsilon_{mn} F^{n-3}_{\mu\nu} + 2 A^{n-3}_{[\mu } K^{\phantom{n-3}}_{\nu ] m n}\\
\ds K_{\mu m n} = b,_{\mu} \epsilon_{mn}

\ea
$$
and, therefore, its non-zero components are given by the expressions
\be\label{K}
\ba{lll}
\ds K_{013} = \frac1{16\pi}\,p^2 (16 b\, e^1 - e^4) \cos\theta, & \ds K_{014} = \frac{e^3}8+2 b \, e^2, & \ds K_{015} = \frac{e^4}8 - 2 b \, e^1,\\
\ds K_{023} = -\frac1{16\pi}\, (16 b\, e^1 p^2 + e^4 p^4) r\sin\theta, & \ds K_{234} = -\frac{b p^2}{2\pi}\sin\theta, & \ds K_{235} = -\frac{p^4}{32\pi} \sin\theta.\\
\ea \ee As it has been mentioned above, the field
strength~$H_\indL{MNK}$ should be eliminated by means of its
equations of motion. Since we perform our calculations up
to~$\alpha{\,}'{}^2$ order, one has to resolve these equations of
motion up to the~$\alpha{\,}'{}^2$ order, what seems to be quite
sophisticated due to the nonlinear nature of the
action~(\ref{MT}). In order to eliminate~$H_\indL{MNK}$, we
perform the following trick~\cite{Prester:2008iu}: we use the
Ansatz (dictated by the symmetry of the problem) for its
non-vanishing components \be \ba{lll}
\ds H_{013} = h_1 \cos\theta, &\ds\quad H_{014} = h_2, &\ds\quad H_{015} = %h_3 =
-\frac{2\pi}{p^2}h_1,\\[0.7em]
\ds H_{023} = h_3 r \sin\theta, &\ds \quad H_{234} = h_4
\sin\theta, &\ds\quad H_{235} = h_5 \sin\theta \ea \ee and include
the fields~$h_i$ in the set of dynamical fields, with respect to
which the entropy function is to be minimised \be\label{extrE}
\ba{ll}
\ds\der{\E}{h_i} = \der{\E}{e^\indU{\L}} =\der{\E}{v_1} = \der{\E}{v_2} =0, &\quad (a) \\[0.75em]
\ds\der{\E}{a}= \der{\E}{b}= \der{\E}{c}= \der{\E}{u_1}= \der{\E}{u_2}= \der{\E}{u_s} = 0. &\quad (b)
\ea
\ee
The entropy function acquires the form
\be
\ba{l}
\ds \E=2 \pi  \left(e^1 q_1 + e^3 q_3 + \frac{u_s v_2}{4} - \frac{u_s v_1}{4} - 16 b h_1 \pi - 32 b h_4 \pi  e^1 - 32 b h_5 \pi  e^2
    - 2 h_5 \pi  e^3 \right.\\
\ds\phantom{\E=}
    \left.+ 2 h_4 \pi  e^4  \frac{a e^4 p^2}{4 \pi } - \frac{h_2 p^4}{2} + \frac{a e^2 p^4}{4 \pi }
    - \frac{h_3^2 u_s}{16 v_2} + \frac{h_3 h_4 e^1 u_s}{4 v_2} - \frac{h_4^2 (e^1)^2 u_s}{4 v_2}
    + \frac{h_3 h_5 e^2 u_s}{4 v_2}\right.\\
\ds\phantom{\E=}
    -\left.\frac{h_4 h_5 e^1 e^2 u_s}{2 v_2} - \frac{h_5^2 (e^2)^2 u_s}{4 v_2} +\frac{h_4^2 u_s v_1}{16 u_1^2 v_2}
    +\frac{c h_4 h_5 u_s v_1}{8 u_1^2 v_2} + \frac{h_5^2 u_s v_1}{16 u_2^2 v_2}  + \frac{c^2 h_5^2 u_s v_1}{16 u_1^2 v_2} \right.\\
\ds\phantom{\E=}
    \left.+ \frac{u_2^2 (p^2)^2 u_s v_1}{64 \pi ^2 v_2}
    - \frac{h_2^2 u_s v_2}{16 u_1^2 v_1} - \frac{u_1^2 e_1^2 u_s v_2}{4 v_1}
    - \frac{c^2 u_2^2 e_1^2 u_s v_2}{4 v_1} + \frac{c u_2^2 e_1 e_2 u_s v_2}{2 v_1} \right.\\
\ds\phantom{\E=} \left.
    - \frac{u_2^2 e_2^2 u_s v_2}{4 v_1} - \frac{c^2 \pi ^2 h_1^2 u_s v_2}{4 u_1^2 (p^2)^2 v_1} %\right.\\
%\ds\phantom{\E=}
    -%\left.
    \frac{\pi ^2 h_1^2 u_s v_2}{4 u_2^2 (p^2)^2 v_1} + \frac{c \pi  h_1 h_2 u_s v_2}{4 u_1^2 {p^2} v_1}\right)
    + O(\alpha{\,}') \;.
\ea \ee Here, for the sake of brevity, we avoid reporting the
explicit expression we derived for the first $\alpha{\,}'$ order
corrections. When minimizing this expression, we perform an
expansion of all dynamical fields over~$\alpha{\,}'$, i.e.
$$ a = a_{(0)}+ \alpha{\,}' a_{(1)} + \ldots, \qquad b = b_{(0)}+ \alpha{\,}' b_{(1)} + \ldots,\qquad\mbox{etc}.$$
In this way one gets ``classical'' solutions to~(\ref{extrE})
\be\label{sol0}
\ba{l}
\ds e^1_{(0)} =  \frac1{4\pi q_1}\,\sqrt{-p^2 p^4 q_1 q_3}, \quad e^2_{(0)} = e^4_{(0)} =  0, \quad
e^3_{(0)} =  \frac1{4 \pi  q_3}\,\sqrt{-p^2 p^4 q_1 q_3}, \\
\ds v_1^{(0)} = v_2^{(0)} = \frac1{{8 \pi ^2}} \left(e^{-{\alpha_2}} + e^{ {\alpha_2}}\right) \,p^2 q_3 \\[0.7em]
\ds h_1^{(0)} = - h_3^{(0)} = -\,\frac{1 - e^{2\alpha_1}}{1 + e^{2\alpha_1}} \frac{p^2 q_3}{4\pi^2},\quad
h_2^{(0)} = -\frac1{2\pi p^4} \sqrt{\strut-p^2 p^4 q_1 q_3},\\
\ds h_4^{(0)} = -\,\frac{1 - e^{2\alpha_1}}{1 + e^{2\alpha_1}} \frac{\sqrt{\strut-p^2 p^4 q_1 q_3}}{2\pi p^4}, \qquad
h_5^{(0)} = \frac{q_3}{2 \pi },\\
\ds a_{(0)} =  4\pi\, \frac{1-e^{2 \alpha_1} }{1 + e^{2 \alpha_1}}\sqrt{-\frac{ q_1 q_3}{p^2 p^4}},\quad
b_{(0)} = -\, \frac1{16}\frac{1-e^{2 \alpha_2} }{1 + e^{2 \alpha_2}}\sqrt{-\frac{p^4 q_1}{p^2 q_3}}, \\
\ds c_{(0)} =  -\,\frac{1-e^{2 \alpha_3} }{1 + e^{2 \alpha_3}} \sqrt{-\frac{ p^2 q_1}{p^4 q_3}},\quad
u_{1}^{(0)} = \sqrt{\frac{4\,e^{\alpha_1+\alpha_3}}{{(1 + e^{2 \alpha_1})(1 + e^{2 \alpha_3})}}}
\sqrt{\rule{0pt}{1.5em}-\frac{q_1}{p^4}}, \\
\ds  u_{2}^{(0)} =
\sqrt{\frac{e^{\alpha_1}(1+e^{2\alpha_3})}{{e^{\alpha_3}(1 + e^{2
\alpha_1})}}} \sqrt{\rule{0pt}{1.5em}\frac{q_3}{p^2}}, \quad
u_s^{(0)} = \frac{16\pi e^{\alpha_2}}{1+e^{2\alpha_2}}
\sqrt{-\frac{p^4 q_1}{p^2 q_3}}. \ea \ee
Here the three real parameters~$\alpha_i$ are restricted by one
condition~\cite{Bellucci:2008sv}
$$ \alpha_1+\alpha_2+\alpha_3=0 ,$$
so that the presence of two independent unconstrained parameters
indicates the presence of two flat directions.

The charges above are connected with standard~$stu$ black hole
charges by the following transformations:
$$
q_1=4 \,\pi \,Q_0,\quad q_3=-\frac{P^1}{2},\quad p^2=\frac{P^3}{2},\quad p^4=- 4\, \pi\, P^2.
$$
One can check this by making a transformation from $SO(2,2)$
%with the "metric" $L=\left( \ba{cc} 0  & I \\ I & 0 \ea \right)$
to~$\left(SU(1,1)/U(1)\right)^3$ basis~\cite{Bellucci:2008sv,Behrndt:1996hu}.
%To get full matching with~\cite{Bellucci:2008sv} one need change moduli as follows
%$$
%a=-4 \,\pi \, x^1,\quad b=\frac{\pi }{2}\, y^1,\quad c= -z^1,\quad
%u1= \sqrt{x^2\, z^2},\quad u2= \sqrt{\frac{x^2} {z^2}},\quad u_s=
%64\, \pi ^2\, y^2
%$$

 Na\"ive substitution of the zero order solution into the entropy function makes it
 dependent on these parameters. This effect appears to contradict the attractor mechanism paradigm.
 However, a subtler analysis shows that these parameters get fixed and the correctness of the
 attractor mechanism is restored on the quantum level too.

To illustrate how the dependence on these parameters drops out, we
present in details the first order correction to the entropy. So,
the ``classical'' solution~(\ref{sol0}) yields the value of the
entropy up to the~$\alpha{\,}'{}^1$ order
$$ \E = \sqrt{-p^2 p^4 q_1 q_3} + 16 \pi^2  \alpha{\,}' e^{\alpha_2}\frac{3 + 2 e^{2 \alpha_2} + 3 e^{4 \alpha_2}}{\left( 1 + e^{2 \alpha_2}\right)^3}
\sqrt{-\frac{p^4 q_1}{p^2 q_3}} + O(\alpha{\,}'{}^2). $$ Let us note
that even including axions the non-BPS solution the first order
correction to the entropy does not contain any contribution form
the Chern--Simons term, in an agreement with~\cite{Sahoo:2006pm}.
Sticking to the first order approximation for the entropy, the
only thing to know about the first order solution to the
eq.~(\ref{extrE}) is that it exists. In general, this is not
guaranteed automatically, since the matrix of the second
derivatives of the entropy function is degenerate. In the case
under consideration the first order solution exists if
$$\alpha_2=0.$$
In this case the first order solution acquires the following form:
$$
\ba{l} \ds h_1^{(1)}= \frac{a_1 {p^2} \sqrt{\strut-{p^2} {p^4} q_1
q_3}}{16 \pi ^3 q_1}, \quad \ds h_2^{(1)}= -\frac{32 e^{2
{\alpha_1}} \pi  q_1}{\left(1+e^{2 {\alpha_1}}\right)^2
\sqrt{\strut-{p^2} {p^4} q_1 q_3}}, \quad
\ds h_5^{(1)}= -\frac{8 \left(-1+e^{2 {\alpha_1}}\right)^2 \pi }{\left(1+e^{2 {\alpha_1}}\right)^2 {p^2}},\\
\ds h_3^{(1)}= \frac{\left(1+e^{2 {\alpha_1}}\right) a_1 (p^2)^2
{p^4} q_3-96 \left(-1+e^{2 {\alpha_1}}\right) \pi^3
    \sqrt{\strut-{p^2} {p^4} q_1 q_3}}{16 \left(1+e^{2 {\alpha_1}}\right) \pi^3 \sqrt{\strut-{p^2} {p^4} q_1 q_3}},\\
\ds h_4^{(1)}= -\frac{\left(1+e^{2 {\alpha_1}}\right) a_1 (p^2)^2
{p^4} q_3+64 \left(-1+e^{2 {\alpha_1}}\right) \pi^3
    \sqrt{\strut-{p^2} {p^4} q_1 q_3}}{8 \left(1+e^{2 {\alpha_1}}\right) \pi^2 {p^2} {p^4} q_3},
\ea
$$
$$
\ba{l} \ds e^1_{(1)}= -\frac{4 \pi  {p^4}}{\sqrt{\strut-{p^2} {p^4}
q_1 q_3}}, \qquad e^3_{(1)}= \frac{4 \pi  {p^4} q_1}{q_3
\sqrt{\strut-{p^2} {p^4} q_1 q_3}}, \qquad e^2_{(1)}= e^4_{(1)}= 0,
\quad v_1^{(1)}=  v_2^{(1)} = 0, \ea
$$
$$
\ba{l} \ds u_1^{(1)} = \frac{e^{-{\alpha_1}}}{8 \left(1+e^{2
{\alpha_1}}\right)^3 \pi  {p^2} (p^4)^2 \sqrt{-\frac{q_1}{{p^4}}}
q_3}
     \left[-128 e^{2 {\alpha_1}} \left(-1+e^{4 {\alpha_1}}\right) \pi  b_1 {p^2} q_3 \sqrt{-{p^2} {p^4} q_1 q_3}+\right.\\
\ds \phantom{u_1^{(1)} =}
    \left.+{p^4} \left(128 e^{2 {\alpha_1}} \left(1+6 e^{2 {\alpha_1}}+e^{4 {\alpha_1}}\right) \pi ^3 q_1+\left(-1+e^{2 {\alpha_1}}\right) \left(1+e^{2 {\alpha_1}}\right)^3 a_1 {p^2} \sqrt{-{p^2} {p^4} q_1 q_3}\right)\right],\\
\ds u_2^{(1)}= \frac{8 \left(-1+e^{2 {\alpha_1}}\right) \left(\left(1+e^{2 {\alpha_1}}\right) b_1 (p^2)^2 q_3^2-\left(-1+e^{2 {\alpha_1}}\right) \pi ^2 \sqrt{-{p^2} {p^4} q_1 q_3}\right)}{\left(1+e^{2 {\alpha_1}}\right)^2 (p^2)^2 \sqrt{\frac{q_3}{{p^2}}} \sqrt{-{p^2} {p^4} q_1 q_3}},\\
\ds c_{(1)} = -\frac{{p^2} q_1}{16 \left(1+e^{2 {\alpha_1}}\right)^3
\pi  \left(-{p^2} {p^4} q_1 q_3\right){}^{3/2}}
    \left[-1024 e^{2 {\alpha_1}} \left(1+e^{2 {\alpha_1}}\right) \pi  b_1 {p^2} q_3 \sqrt{-{p^2} {p^4} q_1 q_3}+\right.\\
    \ds\phantom{c_{(1)} =}
    \left.+{p^4} \left(-1024 e^{2 {\alpha_1}} \left(-1+e^{2 {\alpha_1}}\right) \pi ^3 q_1+4 \left(1+e^{2 {\alpha_1}}\right)^3 a_1 {p^2} \sqrt{-{p^2} {p^4} q_1 q_3}\right)\right],\\
\ds u_s^{(1)} = -\frac{128 \pi ^3 \sqrt{-{p^2} {p^4} q_1
q_3}}{(p^2)^2 q_3^2}, \ea
$$
Let us note that the two scalar fields (in our case these
are~$a_{(1)}$ and~$b_{(1)}$) remain undefined in this
approximation. With vanishing~$\alpha_2$ the first order
correction to the entropy does not depend on any free parameter
anymore and the second order correction acquires a much simpler
form, so that we may write it down as
$$ \E = \sqrt{\strut -p^2 p^4 q_1 q_3} + 16 \pi^2  \alpha{\,}'  \sqrt{-\frac{p^4 q_1}{p^2 q_3}}
    - 16 \pi^4 \alpha{\,}'{}^2 \frac{p^4 q_1}{ p^2 q_3 }
    \frac{1 - 34 e^{\alpha_1} + e^{4\alpha_1} }{\sqrt{\strut-p^2 p^4 q_1 q_3}\left(1 + e^{2\alpha_1} \right)^2}
    + O(\alpha{\,}'{}^3). $$

Performing analogous steps to second order, one can check that
requiring the existence of the second order corrections to the
solution of eqs.~(\ref{extrE}) yields~$\alpha_1=0$. The solution
we obtained in this way, corresponding to $\alpha_i=0$, coincides
with the axion free solution given in
\cite{Prester:2008iu,Sahoo:2006pm}. The only difference is that we
derived this solution keeping the full axion dynamics, without
truncation. It is only on the horizon that the axion contribution
vanishes.

The value of the entropy up to the second order is then given by
\be\label{Entropy2} \E = \sqrt{\strut -p^2 p^4 q_1 q_3} + 16 \pi^2
\alpha{\,}' \sqrt{-\frac{p^4 q_1}{p^2 q_3}}
    - 128 \pi^4 \alpha{\,}'{}^2 \frac{p^4 q_1}{ p^2 q_3 }
    \frac{1}{\sqrt{\strut-p^2 p^4 q_1 q_3}} + O(\alpha{\,}'{}^3).
\ee
One can see that this expression shares the general features pointed out in~(\ref{VPQ}).

\subsection*{Exact solutions}
Despite the cumbersome and entangled structure of the above non-BPS
perturbative solutions to the extremization condition of the entropy
function, one can cast them in exact form\footnote{only
non-vanishing fields are written down\label{footnote}}
\be\label{ourSol} \ba{ll}
%\ds a =  0,\quad b =  0, \quad c =  0, \quad e^2 =  0,
%     \ds \quad e^4 =  0,
\ds v_1 = v_2 =  \frac{p^2 q_3}{4 \pi ^2}, \\[0.5em]
\ds h_2 = - \frac{\sqrt{- p^4 q_1  p^2 q_3}}{2\pi p^4}\, \frac1{\sqrt{1 + \frac{32 \pi ^2 \alpha '}{p^2 q_3}}}, & \quad
\ds h_5 =  \frac{q_3}{2\pi}, \\[0.5em]
\ds e^1 =  \frac{\sqrt{- p^4 q_1  p^2 q_3}}{4\pi q_1}\, \sqrt{1 + \frac{32 \pi ^2 \alpha '}{p^2 q_3}}, & \quad
\ds e^3 =  \frac{\sqrt{- p^4 q_1  p^2 q_3}}{4\pi q_3}\, \frac1{\sqrt{1 + \frac{32 \pi ^2 \alpha '}{p^2 q_3}}}, \\[0.5em]
\ds  u_1 =  \sqrt{-\frac{q_1}{p^4}} \, \frac1{\sqrt{1 + \frac{32 \pi ^2 \alpha '}{p^2 q_3}}},
    \quad u_2 =  \sqrt{\frac{q_3}{p^2}}, &\quad
\ds u_s =  - \frac{8 \pi p^4 q_1}{\sqrt{-p^4 q_1 p^2 q_3}}\frac1{\sqrt{1 + \frac{32 \pi ^2 \alpha '}{p^2 q_3}}},
\ea
\ee
for the charges
\be\label{ourCh}
q_1 < 0, \quad q_3 > 0, \quad p^2 >0, \quad p^4 > 0.
\ee

Once the close form of the solution is obtained, one can see that
the perturbation expansion that we were performing is valid when
$$\frac{32\,\pi^2 \alpha{\,}'}{ |p^2 q_3|} \ll 1.$$
If one fixes the value of~$\alpha{\,}'$, then one may understand this
formula as a condition on the charges
\be\label{p2q3}
 |p^2 q_3| \gg 1,
\ee
when classical effects become dominant. As we will see later,
the condition~(\ref{p2q3}) fails for small black holes.

In the paper~\cite{Prester:2008iu} another exact non-BPS solution
was found, which in our notations reads$^{\ref{footnote}}$
\be\label{presterSol} \ba{ll}
%\ds a =  0,\quad b =  0, \quad c =  0, \quad e^2 =  0, \quad e^4 =  0, \qquad
\ds  v_1 = v_2 =  -\frac{p^2 q_3}{4 \pi ^2}\left( 1 - \frac{32 \pi^2\alpha{\,}'}{p^2 q_3}\right),\\
\ds h_2 = - \frac{\sqrt{\strut- p^4 q_1  p^2 q_3}}{2\pi p^4}\, \sqrt{1 - \frac{32 \pi ^2 \alpha '}{p^2 q_3}}, &\quad
\ds h_5 =  \frac{q_3}{2\pi}\left( 1 - \frac{32 \pi^2\alpha{\,}'}{p^2 q_3}\right), \\
\ds e^1 =  \frac{\sqrt{- p^4 q_1  p^2 q_3}}{4\pi q_1}\, \sqrt{1 - \frac{32 \pi ^2 \alpha '}{p^2 q_3}}, &\quad
\ds e^3 =  \frac{\sqrt{\strut- p^4 q_1  p^2 q_3}}{4\pi q_3}\, \frac1{\ds\sqrt{1 - \frac{32 \pi ^2 \alpha '}{ p^2 q_3}}}, \\
\ds  u_1 =  \sqrt{\frac{q_1}{p^4}},\quad
    u_2 = \sqrt{-\frac{q_3}{p^2}} \, {\sqrt{1 - \frac{\ds 32 \pi ^2 \alpha'}{\ds p^2 q_3}}}, &\quad
\ds u_s =   \frac{8 \pi p^4 q_1}{\sqrt{\strut-p^4 q_1 p^2 q_3}}\frac1{\sqrt{1 - \frac{\ds 32 \pi ^2 \alpha '}{\ds p^2 q_3}}},
\ea
\ee
which is valid for the
following charges: \be\label{presterCh} q_1 <0, \quad q_3 < 0,
\quad p^2 >0, \quad p^4 < 0. \ee
Both solutions correspond to the
same value of the entropy
\be\label{entropyVal}
S=\sqrt{\strut-p^2 p^4 q_1 q_3} \sqrt{1+\frac{32 \pi ^2 \alpha '}{|p^2 q_3|}}. %= \sqrt{\strut |p^2 p^4 q_1 q_3|+32 \pi ^2 \alpha ' |p^4 q_1|}.
\ee In the classical limit~$\alpha{\,}' =  0$ the Hessian matrix of the
solutions~(\ref{ourSol}) and~(\ref{presterSol}) has the following
eigenvalues: \be\ba{l} \ds \lambda_1 = \lambda_2 = 0,\quad
\lambda_3 = 2 |{p^4}| \sqrt{\strut- \frac{{p^2} {p^4} q_3}{q_1}} ,
\quad
\lambda_4 = \frac{|{p^2} q_3|}{64 \pi^2}\sqrt{\strut - \frac{{p^2} q_3}{{p^4} q_1}}, \\[0.9em]
\ds \lambda_5 = 2 {p^2} \sqrt{\strut- \frac{{p^2} {p^4} q_1}{q_3}}
, \quad \lambda_6 = \frac{\ds (p^4)^2 q_3^2 + 256 (p^2)^2 q_3^2
+\frac{(p^4)^2 (p^2)^2}{16\pi^2}  }{\sqrt{\strut-{p^2} {p^4} q_1
q_3}}. \ea \ee When turning on the quantum effects, one finds that
the eigenvalue~$\lambda_1$ remains equal to zero,
while~$\lambda_2$ acquires a positive correction \be \lambda_2 =
32\pi^4 |{p^4}| \sqrt{\strut-\frac{{p^4}}{{p^2} q_1 q_3}}
\frac{(p^4)^2 + 4096 \pi^2 q_3^2}{16 \pi^2 (p^4)^2 q_3^2 + (p^2)^2
(p^4)^2 + 4096 \pi^2 q_3^2 (p^2)^2}\,\alpha{\,}'^2 + O(\alpha{\,}'^3), \ee
so that one can say that the solutions are ``stable''.

To complete the list of solutions, we present as well two
additional solutions. One of them, found in~\cite{Prester:2008iu}, is a BPS solution (see footnote~\ref{footnote} on p.\pageref{footnote})
\be\label{PresterBPS} \ba{ll}
\ds v_1 = v_2 =  -\frac{p^2 q_3}{4 \pi ^2} \left(1-\frac{32 \pi ^2 \alpha '}{p^2 q_3}\right), \\[0.75em]
\ds h_2 =  -\frac{\sqrt{\strut p^2 p^4 q_1 q_3} }{2 \pi  p^4}\frac{1-\frac{32 \pi ^2 \alpha '}{p^2 q_3}}{\sqrt{1-\frac{64 \pi ^2 \alpha '}{p^2 q_3}}},&
    \quad \ds h_5 =  \frac{q_3}{2 \pi } \left(1-\frac{32 \pi ^2 \alpha '}{p^2 q_3}\right),\\[0.75em]
\ds e^1 =  \frac{\sqrt{\strut p^2 p^4 q_1 q_3}}{4 \pi  q_1} \sqrt{1-\frac{64 \pi ^2 \alpha '}{p^2 q_3}},&
    \quad \ds e^3 =  \frac{\sqrt{\strut p^2 p^4 q_1 q_3}}{4 \pi  q_3 } \frac1{\sqrt{1-\frac{ 64 \pi ^2 \alpha '}{p^2 q_3}}},\\[0.75em]
\ds u_1 =  \sqrt{-\frac{q_1}{p^4}\frac{ 1-\frac{32 \pi ^2 \alpha '}{p^2 q_3}}{1-\frac{64 \pi ^2 \alpha '}{p^2 q_3}}}, &
    \quad \ds u_2 =  \sqrt{-\frac{q_3}{p^2}} \sqrt{1-\frac{32 \pi ^2 \alpha '}{p^2 q_3}},
    \quad \ds u_s =  \frac{8 \pi \sqrt{\frac{p^4 q_1}{p^2  q_3}}}{ \sqrt{1-\frac{64 \pi ^2 \alpha '}{p^2 q_3}}}
\ea
\ee
and is valid for the charges
$$ q_1>0,\quad q_3<0,\quad p^2>0,\quad p^4<0. $$
The entropy is then given by
\be
S=\sqrt{\strut p^2 p^4 q_1 q_3} \sqrt{1+\frac{64 \pi ^2 \alpha '}{|p^2 q_3|}}.
\ee
The other solution acquires no~$\alpha{\,}'$ corrections
\be\label{nonBPSZ=0}
\ba{l}
\ds v_1 = v_2 = \frac{ p^2 q_3}{4 \pi ^2}\,, \quad e^1 =  \frac{p^4}{4\pi}
\sqrt{\frac{p^2 q_3}{p^4 q_1}}\,,
\quad e^3 =  \frac{p^2}{4\pi} \sqrt{\frac{p^4 q_1}{p^2 q_3}}\,,\\[0.75em]
\ds h_2 =  -\frac{q_1}{2\pi} \sqrt{\frac{p^2 q_3}{p^4 q_1}}\,, \quad h_5 =  \frac{q_3}{2 \pi }\,,
\quad u_1 =  \sqrt{\frac{q_1}{p^4}}\,,\quad u_2 = \sqrt{\frac{q_3}{p^2}}\,,\quad u_s =  8\pi  \sqrt{\frac{p^4 q_1}{p^2 q_3}}.
\ea
\ee
It is valid for the charges
$$ q_1>0,\quad q_3>0,\quad p^2>0,\quad p^4>0$$
and the corresponding entropy is equal
$$S=\sqrt{\strut p^4 q_1 p^2 q_3}. $$
Despite the fact that it has no quantum corrections, this solution
is non-BPS but with vanishing central charge. These two solutions
are stable in the classical limit, and we suppose that quantum
effects do not spoil this feature.

\subsection*{Small black holes}

Out of the solutions derived in closed form in previous section,
only two admit small black hole limits, i.e. a~$1/2$-BPS~(\ref{PresterBPS}) and a non-BPS~$Z\neq 0$~(\ref{presterSol}), solutions.
The remaining two solutions are non-BPS and do not contain small
black holes (one of them, i.e. the non-BPS $Z=0$ solution, remains
even unaffected by quantum corrections). By definition, a small
black hole has vanishing classical entropy. In the solutions
reported in the previous section, the small black hole limit
corresponds to
\be\label{smallBHLimit}
q_3=0.
\ee
When this limit
is considered, only the two solutions~(\ref{PresterBPS}) and~(\ref{presterSol}) remain regular, while for the remaining two the
radii of the $AdS_2$ and $S^2$ spaces go to zero. Furthermore, the
non-BPS~$Z=0$ solution~(\ref{nonBPSZ=0}) has vanishing entropy,
whereas the non-BPS $Z\neq 0$ solution~(\ref{ourSol}) has the same (non
vanishing) value of the entropy as the solution~(\ref{presterSol}).

One can easily understand that the perturbation theory over the
parameter~$\alpha{\,}'$ fails for small black holes. Indeed the
genuine parameter to make a perturbative expansion is~$q_3$ for
small black holes and~1/$q_3$ for large ones.

Taking the limit~(\ref{smallBHLimit}) in the solutions~(\ref{presterSol}) one
gets the non-BPS small black hole solution
\be\label{smallBHsol}
\ba{l} \ds v_1 =  v_2 =  8 \alpha ', \quad h_2 =  2 \sqrt{\strut
2\alpha{\,}' \, \frac{q_1}{p^4}} ,\quad h_5 = -\frac{16 \pi  \alpha
'}{p^2},\quad e^1 = -\sqrt{\strut 2\alpha{\,}' \frac{p^4}{q_1}},\\
\ds e^3 = -\frac{p^2}{16 \pi^2} \sqrt{\frac{p^4 q_1}{2 \alpha{\,}'}},\quad
\ds u_1 =  \sqrt{\frac{q_1}{{p^4}}},\quad u_2 =  \frac{4 \sqrt{2}
\pi  \sqrt{\alpha '}}{{p^2}}, \quad u_s =  \sqrt{\frac{2 p^4
q_1}{\alpha{\,}' }}\,, \ea \ee
with the corresponding entropy
$$
S=4\pi \sqrt{2\,\alpha{\,}' p^4 q_1}.
$$

Calculating the Hessian matrix on the solution~(\ref{smallBHsol}),
one obtains the following eigenvalues:
\be \ba{l}
\ds \lambda_1 = 0,\quad \lambda_2 = -\frac{(p^2)^4 (p^4)^2+36 ( 8\pi)^6 (p^2)^2
\left(\alpha '\right)^2+ 9 (4 \pi)^6 (p^4)^2 \left(\alpha '\right)^2}
    {512 \sqrt{2} \pi^3 (p^2)^2 \sqrt{\strut {p^4} q_1 \alpha '}}, \\
\ds \lambda_3 = -\frac{16 \pi}3 p^4 \sqrt{\strut 2 \alpha{\,}'\,\frac{p^4}{q_1}}, \quad
\lambda_4 =  -\frac{\sqrt{2}}{\pi} \frac{ (p^2)^2 {p^4} q_1+4
\pi ^2 \alpha '{}^2}{  \sqrt{\strut {p^4} q_1 \alpha '}},\quad
\lambda_5 = 0, \quad \lambda_6 = 0. \ea \ee One sees
that~$\lambda_{2,4}<0$ and~$\lambda_3>0$, which means that the
solution~(\ref{smallBHsol}) is not stable.

Now, taking the limit~(\ref{smallBHLimit}) in the
solutions~(\ref{PresterBPS}) one gets the BPS small black hole
solution
\be\label{smallBHsolBPS}
\ba{l}
\ds v_1 =  v_2 =  8
\alpha ', \quad h_2 =  2 \sqrt{\strut \, -\frac{q_1}{\alpha{\,}' p^4}}, \quad h_5 = -\frac{16 \pi  \alpha '}{p^2},\quad e^1 =
2\sqrt{\strut -\alpha{\,}' \frac{p^4}{q_1}},\\
\ds e^3 = -\frac{p^2}{32 \pi^2} \sqrt{\frac{-p^4 q_1}{ \alpha{\,}'}},\quad
u_1 =  \sqrt{-\frac{q_1}{{2p^4}}},\quad u_2 =  \frac{4
\sqrt{2} \pi  \sqrt{\alpha '}}{{p^2}}, \quad u_s =  \sqrt{-\frac{
p^4 q_1}{\alpha{\,}' }}\,, \ea \ee with the corresponding entropy
$$
S=8\pi \sqrt{-\strut\alpha{\,}' p^4 q_1}.
$$
In this case two of the eigenvalues become zero, three of
them become positive and one becomes negative
$$0,\quad 0,\quad \frac{a_1 + \sqrt{\strut a_1^2 + b_1}}{c_1}, \quad \frac{a_1 - \sqrt{\strut a_1^2 + b_1}}{c_1}, \quad
\frac{a_2 + \sqrt{\strut a_2^2 + b_2}}{c_2}, \quad \frac{a_2 - \sqrt{\strut a_2^2 + b_2}}{c_2}, $$
where
$$\small
\ba{ll}
\ds a_1 = 25  ((p^2)^2 {p^4})^2 + 1145 (4\pi)^6 (256 (p^2)^2 + (p^4)^2 ) \alpha{\,}'{}^2, &%\quad
\ds a_2 =  7 (p^2)^2 {p^4} q_1 + 288 \pi^2 (p^4)^2 \alpha{\,}' - 56 \pi^2 \alpha{\,}'{}^2, \\[0.5em]
\ds b_1 = - 1294 (8\pi)^6 ((p^2)^2 {p^4})^2 (256 (p^2)^2 + (p^4)^2 ) \alpha{\,}'{}^2,  &%\quad
\ds b_2 = 12160 \pi^2 (p^4)^2 \left( - (p^2)^2 {p^4} q_1 + 8 \pi^2 \alpha{\,}'{}^2 \right)\alpha{\,}' ,\\
\ds c_1 = 2588 (2\pi)^3 (p^2)^2 \sqrt{-\strut {p^4} q_1 \alpha{\,}'},&%\quad
\ds c_2 = 38 \pi \sqrt{\strut - {p^4} q_1 \alpha{\,}'}\,.
\ea
$$
We can then conclude that, amazingly enough, for small black
holes, all found solutions~-- BPS and non-BPS~-- are not stable.

\section{Conclusion}
In this paper we considered a correction to the black hole entropy
due to the most general higher order derivative terms in the
Einstein--Maxwell action. We demonstrated that the general form of
the corrections to the entropy is in agreement with previously
found results~\cite{Maldacena}. Provided that the perturbation
expansion over classical solutions is valid, the form of the
correction is completely determined by the classical value of the
entropy~$S_0$ (a homogeneous function of second degree in the
charges) and the classical values of the moduli (a homogeneous
function of zero degree in the charges).

The fact that the subleading corrections are singular in~$S_0$
drops us a hint that small black holes are purely quantum objects.
In fact, the considered example of the~$stu$ black hole
illustrates how small black hole solutions are singular
in~$\alpha{\,}'$, so that the standard perturbation expansion fails
and one should find another small parameter to carry out
perturbation theory ($q_3$ rather than 1/$q_3$). Notice, however,
that the inadequacy of the parameter~$\alpha{\,}'$ as a perturbative
parameter for small black holes is valid in general, not only for
the considered example of the ~$stu$ black hole. The only feature
that will depend on the model is the explicit form of the genuine
perturbation parameter corresponding to the small black hole
limit. It would be interesting to find a model independent
definition of the parameter, generalizing its specific
realization~(\ref{smallBHLimit}) in the model treated in this
paper.

We argued that the attractor mechanism remains valid in the
presence of higher order derivative corrections, despite the fact
that in the presence of flat directions in the classical black
hole potential, the corrections to the entropy apparently might
depend on the undefined moduli. This is made possible by
requiring, to each perturbative order in $\alpha{\,}'$, the
fulfillment of a consistency condition, needed for the existence
of a solution to a degenerate equation system, hence fixing some
of the undefined moduli fields. If not all of the scalars~$\phi_0$
get fixed, then the group symmetry of the higher derivative
corrections is a subgroup of the symmetry of the black hole
potential, hence the higher order corrections will not depend on
these non-fixed scalars.

We established a relation between the black hole potential
approach and the entropy function formalism. The black hole
potential can be deduced from the entropy function by eliminating
``superfluous'' fields from it. In the presence of higher
derivative corrections this procedure might be quite difficult, if
not impossible. Even when deriving the black hole potential proves
difficult, the issue of stability of the solutions can be studied
quite easily, entirely within the entropy function formalism. In
this way the stability of the solutions was studied in this paper.
We believe, however, that the validity of proposed method is quite
general and its applicability goes beyond the considered model,
and even beyond the considered class of systems -- black holes in
four dimensions. Hence, it would be interesting to further apply
the method to study the stability also of extended objects, such
as black strings and black branes, as well as rotating black
objects, considered not only in four but also in higher
dimensions.

In this paper we chose a specific model, i.e. the~$N=2$~$d=4$
supergravity model with~$stu$ prepotential, since it is known to
possess flat directions in the non-BPS
branch~\cite{Ferrara:2007tu}. In order to derive a form of the
corrections pertinent to this theory, we started from the action
of heterotic string theory~\cite{Metsaev:1987zx} with
all~$\alpha{\,}'$ corrections included~\cite{Prester:2008iu}.
Consequent compactification down to four dimensions reproduces
a~$stu$ model with all necessary corrections included. The most
interesting features for us reside in this case in the sector of
the axion fields; therefore we did not make any truncation over
the axions.

Apart from the known solutions of the obtained model, we found two
more non-BPS ones. It is noteworthy that one of these solutions
acquires no~$\alpha{\,}'$ corrections and it is a non-BPS solution
with zero central charge.

We investigated the small black hole limit of the above mentioned
solutions and found that the solutions found by us are singular in
such a limit, while the previously known solutions become unstable
in the sense that the corresponding Hessian matrix acquires
positive, negative and zero eigenvalues. It is of interest to investigate whether this property of small black hole solutions is general
or just a peculiarity in the considered~$stu$ model.

\section*{Acknowledgements}
The work of S.B., S. F. and A. Y. have been supported in part by
the European Research Council grant n.~226455, ``SUPERSYMMETRY,
QUANTUM GRAVITY AND GAUGE FIELDS (SUPERFIELDS)''.

The work of A.S. has been supported by a Junior Grant of the
Enrico Fermi Centre, Rome, in association with INFN Frascati
National Laboratories.

A. Y. would like to thank the Department of Physics, Theory Unit
Group at CERN, where this project was completed, for kind
hospitality and stimulating environment.

\end{document}